\documentclass[aps,prl,twocolumn,superscriptaddress,showpacs,preprintnumbers,amsmath,amssymb]{revtex4-2}
\UseRawInputEncoding 

\usepackage{graphicx} 
\usepackage{dcolumn}  
\usepackage{lineno}
\usepackage{color}
\usepackage[dvipsnames]{xcolor}
\usepackage{comment}
\usepackage{ulem}
\usepackage{orcidlink}

\def\ra{\!\rightarrow\!}

\def\simge{\mathrel{%
   \rlap{\raise 0.511ex \hbox{$>$}}{\lower 0.511ex \hbox{$\sim$}}}}
\def\simle{\mathrel{
   \rlap{\raise 0.511ex \hbox{$<$}}{\lower 0.511ex \hbox{$\sim$}}}}

\def\dsp{D^+_s}
\def\Dsphipi{D^+_s\ra\phi\pi^+}
\def\phikk{\phi\ra K^+K^-} 

\def\sigmat{\sigma^{}_t}

\def\trec {t}

\graphicspath{{ps}}

\begin{document}

\vspace*{-3\baselineskip}
\resizebox{!}{3cm}{\includegraphics{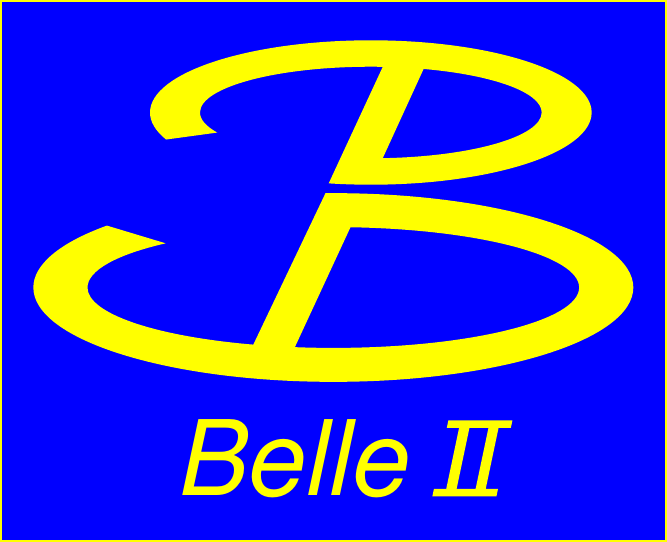}}

\vspace*{\baselineskip}
\begin{flushright}
\end{flushright}

\preprint{Belle Preprint 2023-007}
\preprint{KEK Preprint 2023-5}

\title{
{\boldmath Precise measurement of the $D^+_s$ lifetime at Belle~II }
}

  \author{I.~Adachi\,\orcidlink{0000-0003-2287-0173}} 
  \author{L.~Aggarwal\,\orcidlink{0000-0002-0909-7537}} 
  \author{H.~Aihara\,\orcidlink{0000-0002-1907-5964}} 
  \author{N.~Akopov\,\orcidlink{0000-0002-4425-2096}} 
  \author{A.~Aloisio\,\orcidlink{0000-0002-3883-6693}} 
  \author{N.~Anh~Ky\,\orcidlink{0000-0003-0471-197X}} 
  \author{D.~M.~Asner\,\orcidlink{0000-0002-1586-5790}} 
  \author{H.~Atmacan\,\orcidlink{0000-0003-2435-501X}} 
  \author{T.~Aushev\,\orcidlink{0000-0002-6347-7055}} 
  \author{V.~Aushev\,\orcidlink{0000-0002-8588-5308}} 
% ...
\collaboration{The Belle II Collaboration}

\begin{abstract}
We measure the lifetime of the $D_s^+$ meson using a data sample of 
207~fb$^{-1}$ collected by the Belle~II experiment running at the
SuperKEKB asymmetric-energy $e^+ e^-$ collider. The lifetime is 
determined by fitting the decay-time distribution of a sample 
of $116\times 10^3$ $D_s^+\ra\phi\pi^+$ decays. Our result is 
$\tau^{}_{D^+_s} = (499.5\pm 1.7\,\pm 0.9)$~fs, 
where the first uncertainty is statistical and the second is systematic.
This result is significantly more precise than previous measurements.
\end{abstract}

\maketitle

The lifetime of a particle, like its mass and spin, is one of the fundamental
properties that distinguishes it from other particles. The lifetime is the 
reciprocal of the total decay width, which is the sum of all partial decay 
widths. Each partial width is proportional to the magnitude squared of 
the sum of all decay amplitudes to a final state, and thus every decay 
amplitude potentially affects the lifetime. As a result, the lifetime
can provide information about amplitudes that are difficult to measure or calculate. 

\begin{figure}[ht]
\hbox{\hskip-0.80in
\vbox{
    \includegraphics[width=0.22\textwidth]{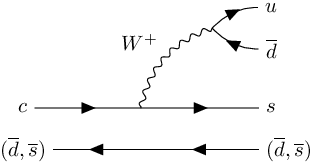}
\vskip0.30in
    \includegraphics[width=0.19\textwidth]{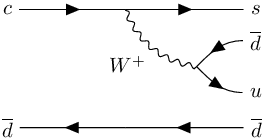}
}
\hskip-1.8in
\vbox{
    \includegraphics[width=0.15\textwidth]{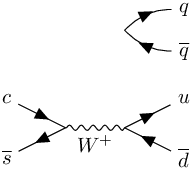}
\vskip0.5in
}
}
    \caption{Left: spectator amplitude (top) and color-suppressed amplitude (bottom). 
            Right: annihilation amplitude.}
    \label{fig:Dfeynman}
\end{figure}

Lifetimes of $D$ mesons are dominated
by partial widths to hadronic final states.
The relatively long lifetime of the $D^+$ meson, 
2.5 times that of the $D^0$, 
implies there is a reduction in hadronic partial widths.
This reduction is attributed to
destructive interference between a ``spectator'' amplitude 
and a color-suppressed amplitude
(Fig.~\ref{fig:Dfeynman}, left)~\cite{Morrison:1989xq}. The 
small difference in lifetimes of the $D^0$ and $D^+_s$ mesons 
is attributed to the dominance of the spectator amplitude for 
hadronic decays and different color factors that enter subdominant 
``exchange'' ($D^0$) and ``annihilation'' ($D^+_s$) amplitudes~\cite{Browder:1996af}.
The latter amplitude for $D^+_s$ decays (Fig.~\ref{fig:Dfeynman}, right)
is Cabibbo-favored and thus plays a larger role than it does for $D^+$ 
decays, in which it is Cabibbo-suppressed.

Hadron lifetimes are difficult to calculate theoretically, as they depend on nonperturbative effects arising from quantum chromodynamics
(QCD). Thus, lifetime calculations are performed using phenomenological
methods such as the heavy quark 
expansion~\cite{Lenz:2014jha,Neubert:1997gu,Uraltsev:2000qw,PhysRevD.88.034004,Kirk:2017juj,Gratrex:2022xpm}.
Comparing calculated values with measured values improves our understanding of 
QCD, which leads to improved 
QCD calculations of other quantities such as hadron masses,
structure functions, etc.~\cite{FlavourLatticeAveragingGroupFLAG:2021npn}.
Measurements of the $D_s^+$ lifetime have been reported by many
experiments~\cite{LHCb:2017knt,FOCUS:2005gui,SELEX:2001miq,E791:1998zjs,CLEO:1999xvl,E687:1993lxk,TaggedPhotonSpectrometer:1987poq};
the world average value is 
$\tau^{}_{D^+_s} = (504\pm 4)$~fs~\cite{ParticleDataGroup:2022pth}.
In this Letter, we present a new measurement of the $D^+_s$ lifetime using 
$\Dsphipi$ decays~\cite{charge-conjugates} reconstructed in 207~fb$^{-1}$ 
of data collected by the Belle~II experiment~\cite{Abe:2010gxa,Kou:2018nap}.
The data were recorded at an $e^+e^-$  center-of-mass energy corresponding 
to the $\Upsilon(4S)$ resonance, and at an energy slightly below.
Our result has significantly greater precision than the world
average value. 

The Belle~II experiment runs 
at the SuperKEKB $e^{+} e^{-}$ collider~\cite{Akai:2018mbz}. The overall 
detector~\cite{Abe:2010gxa} has a cylindrical geometry and includes a two-layer 
silicon-pixel detector (PXD) surrounded by a four-layer double-sided 
silicon-strip detector (SVD)~\cite{Belle-IISVD:2022upf} and a 56-layer 
central drift chamber (CDC). 
These detectors reconstruct tracks (trajectories of charged particles).
Only one sixth of the second layer of the PXD was installed for the data analyzed here.
The axis of symmetry of these detectors, defined as the $z$ axis,
is almost coincident with the direction of the electron beam.
Surrounding the CDC is a time-of-propagation counter (TOP)~\cite{Wang:2017ajq} 
in the central region, and an aerogel-based ring-imaging Cherenkov 
counter (ARICH) in the forward region.
These detectors provide charged-particle identification.
Surrounding the TOP and ARICH is an electromagnetic calorimeter based on 
CsI(Tl) crystals that provides energy and timing measurements for photons 
and electrons. Outside of the calorimeter is an iron flux return for a 
superconducting solenoid magnet. The flux return is instrumented with 
resistive plate chambers and plastic scintillator modules to detect 
muons, $K^0_L$ mesons, and neutrons. The solenoid magnet provides 
a 1.5~T magnetic field that is parallel to the $z$ axis. 

We use Monte Carlo (MC) simulated events to optimize event selection criteria, 
calculate reconstruction efficiencies, and study sources of background. We generate 
$e^+e^-\ra q\bar{q}~(q=u,d,s,c,b)$ events using the {\sc KKMC} package~\cite{Jadach:1999vf} 
and simulate quark hadronization
using the {\sc Pythia 8} package~\cite{Sjostrand:2014zea}. 
Hadron decays are simulated using {\sc EvtGen}~\cite{Lange:2001uf}, and
the detector response is simulated using {\sc Geant4}~\cite{Agostinelli:2002hh}. 
Final-state radiation is included in the simulation via 
{\sc Photos}~\cite{Barberio:1993qi}. Both MC-simulated events and 
collision data are reconstructed using the Belle~II analysis software 
framework~\cite{Kuhr:2018lps,basf2-zenodo}. To avoid introducing bias 
in our analysis, we analyze the data in a ``blind" manner, i.e, we 
finalize all selection criteria and the fitting procedure before 
evaluating the lifetime of signal candidates.

We reconstruct $\Dsphipi$ decays by first reconstructing $\phi\ra K^+K^-$ 
decays and subsequently pairing the $\phi$ candidate with a $\pi^+$ track. We 
select well-measured tracks by requiring that each track have at least one hit 
(measured point) in the PXD, four hits in the SVD, and 30 hits in the CDC.
We select tracks that originate from near the interaction point (IP) by 
requiring $|\delta z|< 2.0$~cm and $\delta r< 0.5$~cm, where 
$\delta z$ is the displacement of the track from the IP
along the $z$ axis, 
and $\delta r$ is the radial displacement in the plane transverse 
to the $z$ axis. The IP position is measured at regular intervals 
of data-taking using $e^+e^-\ra\mu^+\mu^-$ events. The spread of
the IP position is typically $250~\mu$m in the $z$ direction, 
$10~\mu$m in the transverse horizontal direction ($x$), and 
only $0.3$~$\mu$m in the transverse vertical direction ($y$). 
We have checked that none of the above requirements, nor any 
subsequent selection requirements, bias the lifetime measurement.

We identify tracks as pions or kaons based 
on Cherenkov light recorded in the TOP
and ARICH, and specific ionization ($dE/dx$)
information from the CDC and SVD. This information is combined 
to calculate a likelihood ${\cal L}^{}_{K,\pi}$ 
for a track to be a $K^+$ or $\pi^+$. Tracks having a ratio 
${\cal L}_{K}/({\cal L}_{K}+{\cal L}_{\pi})>0.60$ 
are identified as kaon candidates, while tracks having
${\cal L}_{K}/({\cal L}_{K}+{\cal L}_{\pi})<0.55$ 
are identified as pion candidates. These requirements 
are 90\% and 95\% efficient for kaons and pions, respectively.

To reconstruct $\phikk$ decays, we combine two kaon candidate 
tracks having opposite charge and an invariant mass satisfying 
$1.010~{\rm GeV}/c^2 < M(K^+K^-) < 1.030~{\rm GeV}/c^2$. 
This selected range retains 91\% of $\phi\ra K^+K^-$ decays.
We pair $\phi$ candidates with $\pi^+$ tracks to form $D_s^+$
candidates and require that the invariant mass satisfy a loose 
requirement of $1.922~{\rm GeV}/c^2 < M(\phi\pi^+) <2.020~{\rm GeV}/c^2$.
We fit the three tracks to a common vertex using the 
{\sc TreeFitter} algorithm~\cite{Belle-IIanalysissoftwareGroup:2019dlq}. 
The vertex position resulting from the fit is taken as the decay vertex 
of the $D_s^+$. The fit includes a constraint that the $D_s^+$ trajectory 
be consistent with originating from the IP; this constraint improves the 
resolution on the $D_s^+$ decay time by a factor of three.

To eliminate $D_s^+$ mesons originating from $B$ decays, which would
not have a properly determined decay time, we require that the momentum of 
the $D_s^+$ in the $e^+e^-$ center-of-mass frame be greater than 2.5~GeV/$c$. 
This selection eliminates all $D_s^+$ mesons from $B$ decays while retaining
67\% of those produced via $e^+e^-\ra c\bar{c}$.
We reduce background arising from random combinations of $\phi$ and $\pi^+$ 
candidates by requiring $|\cos\theta_K| >0.45$, where 
$\theta_K$ is the angle in the $\phi$ rest frame between the $K^-$ 
momentum and the direction of the $D_s^+$. This requirement reduces 
combinatorial background by~40\% while retaining 90\% of signal decays. 
After applying all selection criteria, about 2\% of events have more 
than one $\Dsphipi$ candidate. 
False signal candidates arise mainly from combinations of $\phi$ decays  
with unrelated $\pi^+$ tracks. These do not peak in $M(\phi\pi^+)$ and are 
counted as background in our fits for signal yield and $D_s^+$ lifetime;
consequently, they have a negligible effect on the fitted lifetime. We thus 
retain all signal candidates.

\begin{figure}[ht]
    \centering
    \includegraphics[width=0.50\textwidth]{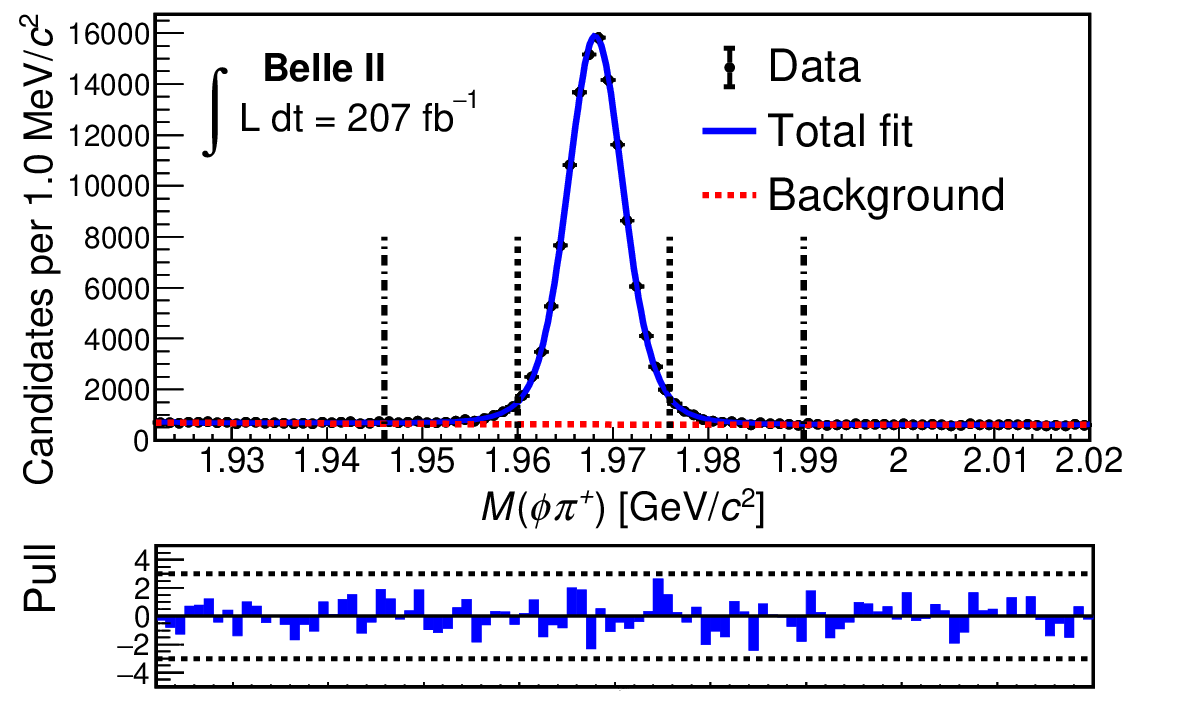}
    \caption{Distribution of $M(\phi\pi^+)$ for $\Dsphipi$ candidates,
        with the fit result overlaid. Black dots correspond to the data; 
       the red dashed curve shows the background component; and the blue 
       solid curve shows the overall fit result.
       Vertical dotted lines denote the signal region, and 
      vertical dot-dashed lines denote the upper and lower boundaries 
      of the lower and upper sidebands (see text). 
       The corresponding pull distribution is shown 
     in the lower panel, where the pull is defined as 
     $({\rm data}-{\rm fit})/({\rm statistical\ uncertainty\ in\ data})$. }
    \label{fig:Mkkpi_plot}
\end{figure}

The final $M(\phi\pi^+)$ distribution is shown in Fig.~\ref{fig:Mkkpi_plot}.
We perform an unbinned maximum likelihood fit to $M(\phi\pi^+)$ to determine
the yield of $\Dsphipi$ decays. The signal shape is modeled as the sum of 
two Gaussian functions and an asymmetric Student's~t distribution.
The background contains no peaking structure ($>\!95\%$ consists of 
random combinations of $\phi$ and $\pi^+$ candidates) and is well-modeled 
by a second-order polynomial. To measure the $D_s^+$ lifetime, 
we select candidates having an invariant mass satisfying
$1.960~{\rm GeV}/c^2 < M(\phi\pi^+) < 1.976~{\rm GeV}/c^2$. 
This range retains 95\% of $\Dsphipi$ decays.
In this signal region, the fit yields 115560 signal decays and 9970
background events; the signal purity
(ratio of signal over the total) is~92\%.

The decay time of a $D_s^+$ candidate is calculated as
\begin{eqnarray}
\trec & = & \left(\frac{\vec{d}\cdot\vec{p}}{p^2}\right) m^{}_{D_s^+} \,,
\label{eqn:decay_time}
\end{eqnarray}
where $\vec{d}$ is the displacement vector from the IP to the $D_s^+$
decay vertex, $\vec{p}$ is the $D_s^+$ momentum, and $m^{}_{D_s^+}$ is the
known $D_s^+$ mass~\cite{ParticleDataGroup:2022pth}.
The average resolution on $\trec$ is 108~fs.
We determine the $D^+_s$ lifetime by performing an unbinned maximum 
likelihood fit to two observables: the decay time $\trec$ and the 
per-candidate uncertainty on $\trec$ ($\sigmat$)
as calculated from the uncertainties on $\vec{d}$ and~$\vec{p}$.
The likelihood function for the $i$th candidate is given by
\begin{eqnarray} \label{eqn:likelihood_lftm}
{\cal L}(\tau | t^i,\sigmat{}\!\!^i) & = & 
f_{\rm sig}\,P_{\rm{sig}}(t^i | \tau, \sigmat{}\!\!^i)\,P_{\rm{sig}}(\sigmat{}\!\!^i)\ +\   
\nonumber \\
 & & \hskip 0.08in
(1-f_{\rm{sig}})\,P_{\rm{bkg}}(t^i|\sigmat{}\!\!^i)\,P_{\rm{bkg}}(\sigmat{}\!\!^i) ,
\end{eqnarray}
where $f^{}_{\rm sig}$ is the fraction of events that are signal $\Dsphipi$ decays;
$P_{\rm sig}(t | \tau,\,\sigmat)$ and $P_{\rm bkg}(t | \sigmat)$ are 
probability density functions (PDFs) for signal and background events
for a reconstructed decay time $\trec$, given a $\dsp$ lifetime $\tau$ 
for signal and an uncertainty $\sigmat$; and $P_{\rm sig}(\sigmat)$ and 
$P_{\rm bkg}(\sigmat)$ are the respective PDFs for~$\sigmat$. To reduce 
highly mismeasured events that are difficult to simulate, we impose loose 
requirements $-2000~{\rm fs}< \trec <4000$~fs and $\sigma^{}_t<900$~fs.
These requirements reject less than 0.1\% of signal candidates.

The signal PDF is the convolution of an exponential function
and a resolution function $R$:
\begin{eqnarray}
P^{}_{\rm sig}(t^i|\tau, \sigmat{}\!\!^i)  
& = & \frac{1}{\tau} \int e^{-t'/\tau}\,R(t^i-t' ; \mu, s, \sigmat{}\!\!^i)\,dt' ,
\label{lftm_pdf}
\end{eqnarray}
where $R(t^i-t' ; \mu, s, \sigmat{}\!\!^i)$ is a single Gaussian function
with mean $\mu$ and a per-candidate standard deviation $s\times \sigmat{}\!\!^i$. 
The scaling factor $s$ accounts for under- or over-estimation of the 
uncertainty $\sigmat{}\!\!^i$. The PDF $P_{\rm bkg}(t\,|\sigmat)$ is determined 
by fitting the decay-time distribution of events in the 
$M(\phi\pi^+)$ ``upper'' sideband 
$1.990~{\rm GeV}/c^2 < M(\phi\pi^+) <2.020~{\rm GeV}/c^2$,
which has no contamination from signal decays with final-state radiation.
We model $P_{\rm bkg}(t|\sigmat)$ as the sum of three asymmetric Gaussians with a common mean. 
We use MC simulation to verify that the decay-time distribution of background events in this
sideband describes well the decay-time distribution of background events in the signal region.

The PDFs $P^{}_{\rm sig}(\sigmat)$ and $P^{}_{\rm bkg}(\sigmat)$ 
are taken to be finely binned histograms. 
The former is determined from the $\sigmat$ distribution of events in the signal region, 
after subtracting the $\sigmat$ distribution of events in the $M(\phi\pi^+)$ sideband. 
The resulting $P^{}_{\rm sig}(\sigmat)$ distribution matches well that of MC-simulated 
signal decays. The $P^{}_{\rm bkg}(\sigmat)$ distribution
is determined from background events in the $M(\phi\pi^+)$ sideband.  
The signal fraction $f_{\rm{sig}}$ is obtained from the earlier fit to the 
$M(\phi\pi^+)$ distribution (Fig.~\ref{fig:Mkkpi_plot}) and fixed in this fit.
Thus, there are three floated parameters: the lifetime $\tau$, and the 
mean parameter $\mu$ and scaling factor $s$ of the resolution function. 
These are determined by maximizing the total log-likelihood
$\sum_i \ln {\cal L}(\tau | t^i,\sigmat{}\!\!^i)$, 
where the sum runs over all events in the signal region.

\begin{figure}[ht]
    \centering
    \includegraphics[scale=0.48]{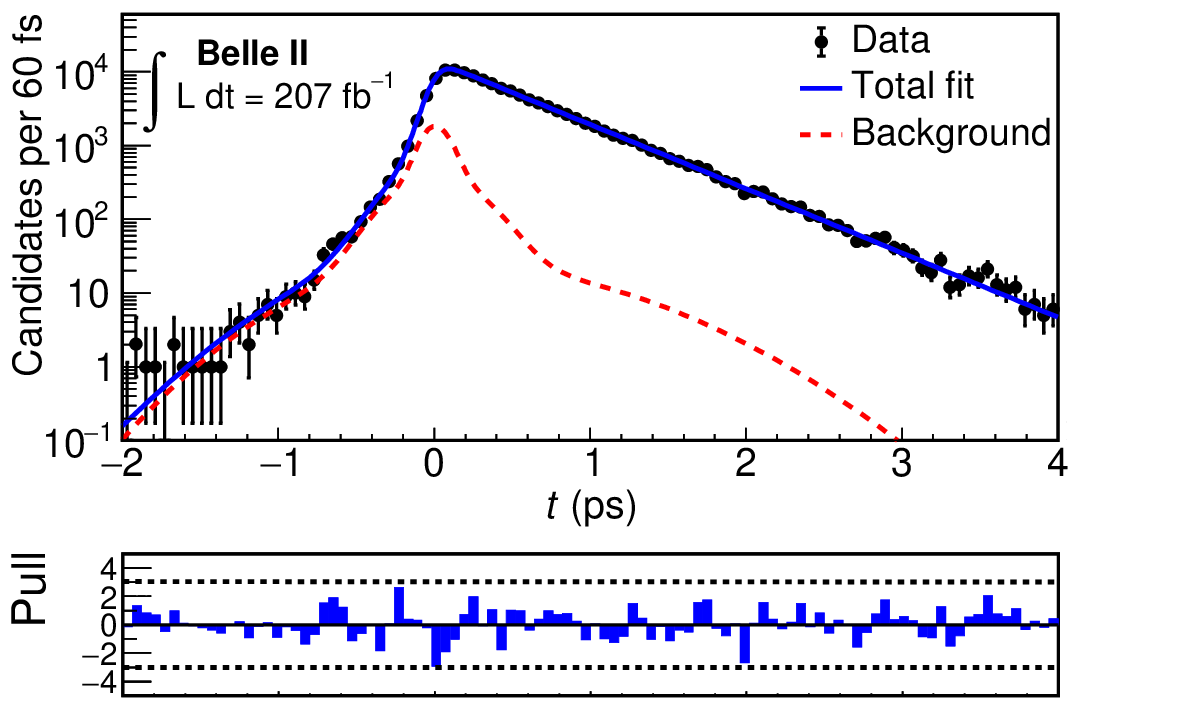}
    \caption{Distribution of $\trec$ for $\Dsphipi$ candidates,
        with the fit result overlaid. Black dots correspond to the data; 
       the red dashed curve shows the background component; and the blue solid 
       curve shows the overall fit result. The corresponding pull distribution 
       is shown in the lower panel.}
    \label{fig:datafit_tproj}
\end{figure}

The result of the fit 
is $\tau = 498.70 \pm 1.71$~fs, where the uncertainty is statistical only. 
The projection of the fit for $\trec$ is shown in Fig.~\ref{fig:datafit_tproj} 
along with the resulting pulls; the $\chi^2$ divided by the number of degrees 
of freedom ($100-4 = 96$) is 1.02.
The values $\mu = 0.56 \pm 0.86$~fs and $s=1.22 \pm 0.01$
obtained for the resolution function are similar to those 
obtained from MC-simulated samples.

The main systematic uncertainties are listed in 
Table~\ref{tab:syst_summary} and evaluated as follows.
Uncertainty arising
from possible mismodeling of the detector response and 
possible correlations between $\trec$ and $\sigmat$ not 
accounted for by the resolution function is assessed
by fitting a large ensemble of MC signal events. 
The mean of the fitted lifetime values
is calculated, and the difference of $-0.85$~fs between 
the mean value and the input value is used to correct the 
fitted lifetime. We assign half of this correction as 
a systematic uncertainty.

There is uncertainty arising from modeling the background decay-time distribution.
We model this distribution using background events in the upper $M(\phi\pi^+)$ sideband 
$1.990~{\rm GeV}/c^2 < M(\phi\pi^+) <2.020~{\rm GeV}/c^2$. To evaluate uncertainty in this model,
we choose a lower sideband $1.922~{\rm GeV}/c^2 < M(\phi\pi^+) <1.946~{\rm GeV}/c^2$, a 
combination of the two sidebands, and also the MC-simulated background distribution
in the signal region. The largest difference observed between the resulting fitted 
lifetime and our nominal result is assigned as a systematic uncertainty. 

We model both signal and background $\sigmat$ distributions using histogram PDFs, 
and there is systematic uncertainty arising from our choice for the number of bins
(i.e., statistical fluctuations of the sideband data used to obtain the histogram PDF).
We evaluate this by changing the number of bins from the nominal value (80) to 
other values in the range 60--400. For each choice of binning, we refit 
for $\tau$. The largest difference observed between the resulting values and 
our nominal value is taken as a systematic uncertainty. 

As measuring the decay time depends on a precise determination of the
displacement vector $\vec{d}$ and momentum $\vec{p}$ (Eq.~\ref{eqn:decay_time}), 
there is uncertainty arising from possible misalignments of the PXD, SVD, and 
CDC detectors. We study the effect of such misalignment using MC events 
reconstructed with various misalignments. Each sample is equivalent in size to that 
of the collision data used. The difference between the fitted value of $\tau$ and 
the result obtained with no misalignment is recorded, and the root-mean-square (r.m.s.) 
of the distribution of differences is taken as the systematic uncertainty due to 
possible detector misalignment. 

There is uncertainty arising from the fraction of signal candidates ($f^{}_{\rm sig}$),
which is fixed in the decay-time fit to the value obtained
from the fit to the $M(\phi\pi^+)$ distribution. We vary this parameter by 
its uncertainty and take the resulting change in the fitted lifetime as a 
systematic uncertainty. 

There is an uncertainty arising from the global momentum scale of the detector,
which is calibrated using the peak position of $D^0\ra K^-\pi^+$ decays.
We evaluate this by varying the global scale factor by its uncertainty 
($\pm 0.06\%$) and assigning the resulting variation in the fitted 
lifetime as a systematic uncertainty.

Finally, we include a systematic uncertainty due to uncertainty in the $D_s^+$ 
mass~\cite{ParticleDataGroup:2022pth}, which is used to calculate the decay 
time from the vectors $\vec{d}$ and $\vec{p}$ [see Eq.~(\ref{eqn:decay_time})].
The total systematic uncertainty is obtained by adding together all individual 
contributions (listed in Table~\ref{tab:syst_summary}) in quadrature. 
The result is $\pm 0.87$~fs.

\begin{table}[ht]
\renewcommand{\arraystretch}{1.2}
\begin{tabular}{lc} 
\hline \hline
Source                                   & Uncertainty (fs)   \\ 
\hline    
Resolution function                      &   $\pm 0.43$  \\
Background $(\trec,\sigma^{}_t)$ distribution  &  $\pm 0.40$       \\
Binning of $\sigmat$ histogram PDF      &   $\pm 0.10$       \\
Imperfect detector alignment             &  $\pm 0.56$       \\
Sample purity                            &   $\pm 0.09$       \\
Momentum scale factor                    &   $\pm 0.28$       \\
$D^+_s$ mass                             &   $\pm 0.02$       \\ \hline
Total                                    &   $\pm 0.87$       \\ 
\hline \hline 
\end{tabular}
\caption{Summary of systematic uncertainties.}
\label{tab:syst_summary}
\end{table}

\begin{figure}[ht]
    \centering
    \includegraphics[scale=0.41]{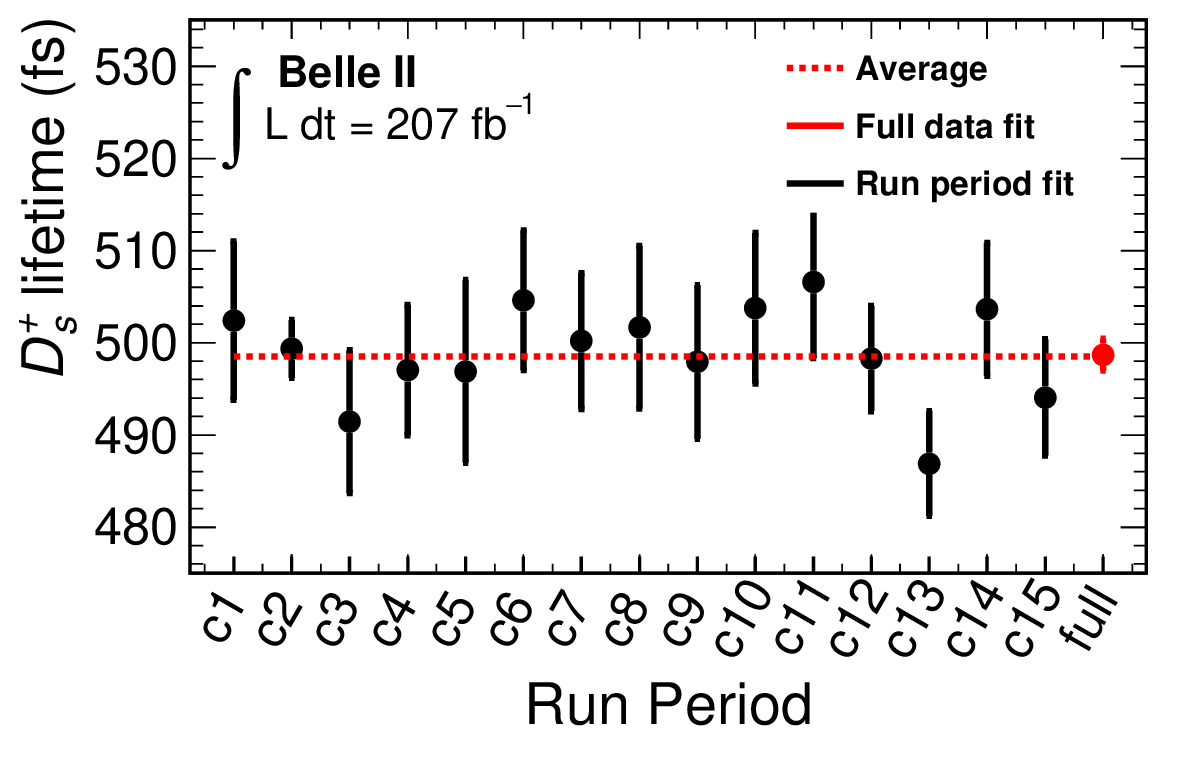}
    \caption{
    Fitted lifetime for different run periods (c1--c15). 
    For these fits, the parameters of the resolution
    function are fixed to the overall fitted values.
    All values are consistent with the overall result, which is plotted 
    as a red data point. The red dashed line shows the average of the
    lifetime results for the different run periods.
    }
    \label{fig:run_period}
\end{figure}

As a final check of our analysis procedure, we divide the data 
sample into subsets based on $D_s^+$ (or $D_s^-$) charge, 
$D_s^+$ momentum, $D_s^+$ polar angle, $D_s^+$ azimuthal 
angle, and data-collection (run) period, and we measure the 
lifetime separately for each subset. All measured values are 
consistent with statistical fluctuations about the overall result.
The fitted lifetime for different run periods is plotted in Fig.~\ref{fig:run_period}.

In summary, we have used $116\times 10^3$ $D_s^+\ra\phi\pi^+$ decays 
reconstructed in 207~fb$^{-1}$ of data recorded by Belle~II 
in $e^+e^-$ collisions at or near the $\Upsilon(4S)$ resonance 
to measure the $D_s^+$ lifetime. The result is
\begin{eqnarray}
\tau^{}_{D^+_s} & = & (499.5 \pm 1.7\,\pm 0.9)~{\rm fs,}
\end{eqnarray}
where the first uncertainty is statistical and the second is systematic. 
This is the most precise measurement to date. It is consistent with, but 
has half the uncertainty of, the current world-average value of 
$(504\pm 4)$~fs~\cite{ParticleDataGroup:2022pth}. It is also consistent 
with theory predictions~\cite{Lenz:2014jha, PhysRevD.88.034004,Gratrex:2022xpm}.
The high precision results from significantly improved decay-time resolution 
as compared to previous $e^+e^-$ experiments. This is due to the smaller beam 
sizes of SuperKEKB, which reduce uncertainty on the IP, and to the smaller 
radius of the first layer of the vertex detector.

This work, based on data collected using the Belle II detector, which was built and commissioned prior to March 2019, was supported by
Science Committee of the Republic of Armenia Grant No.~20TTCG-1C010;
Australian Research Council and research Grants
No.~DP200101792, 
No.~DP210101900, 
No.~DP210102831, 
No.~DE220100462, 
No.~LE210100098, 
and
No.~LE230100085; 
Austrian Federal Ministry of Education, Science and Research,
Austrian Science Fund
No.~P~31361-N36
and
No.~J4625-N,
and
Horizon 2020 ERC Starting Grant No.~947006 ``InterLeptons'';
Natural Sciences and Engineering Research Council of Canada, Compute Canada and CANARIE;
National Key R\&D Program of China under Contract No.~2022YFA1601903,
National Natural Science Foundation of China and research Grants
No.~11575017,
No.~11761141009,
No.~11705209,
No.~11975076,
No.~12135005,
No.~12150004,
No.~12161141008,
and
No.~12175041,
and Shandong Provincial Natural Science Foundation Project~ZR2022JQ02;
the Ministry of Education, Youth, and Sports of the Czech Republic under Contract No.~LTT17020 and
Charles University Grant No.~SVV 260448 and
the Czech Science Foundation Grant No.~22-18469S;
European Research Council, Seventh Framework PIEF-GA-2013-622527,
Horizon 2020 ERC-Advanced Grants No.~267104 and No.~884719,
Horizon 2020 ERC-Consolidator Grant No.~819127,
Horizon 2020 Marie Sklodowska-Curie Grant Agreement No.~700525 "NIOBE"
and
No.~101026516,
and
Horizon 2020 Marie Sklodowska-Curie RISE project JENNIFER2 Grant Agreement No.~822070 (European grants);
L'Institut National de Physique Nucl\'{e}aire et de Physique des Particules (IN2P3) du CNRS (France);
BMBF, DFG, HGF, MPG, and AvH Foundation (Germany);
Department of Atomic Energy under Project Identification No.~RTI 4002 and Department of Science and Technology (India);
Israel Science Foundation Grant No.~2476/17,
U.S.-Israel Binational Science Foundation Grant No.~2016113, and
Israel Ministry of Science Grant No.~3-16543;
Istituto Nazionale di Fisica Nucleare and the research grants BELLE2;
Japan Society for the Promotion of Science, Grant-in-Aid for Scientific Research Grants
No.~16H03968,
No.~16H03993,
No.~16H06492,
No.~16K05323,
No.~17H01133,
No.~17H05405,
No.~18K03621,
No.~18H03710,
No.~18H05226,
No.~19H00682, 
No.~22H00144,
No.~26220706,
and
No.~26400255,
the National Institute of Informatics, and Science Information NETwork 5 (SINET5), 
and
the Ministry of Education, Culture, Sports, Science, and Technology (MEXT) of Japan;  
National Research Foundation (NRF) of Korea Grants
No.~2016R1\-D1A1B\-02012900,
No.~2018R1\-A2B\-3003643,
No.~2018R1\-A6A1A\-06024970,
No.~2018R1\-D1A1B\-07047294,
No.~2019R1\-I1A3A\-01058933,
No.~2022R1\-A2C\-1003993,
and
No.~RS-2022-00197659,
Radiation Science Research Institute,
Foreign Large-size Research Facility Application Supporting project,
the Global Science Experimental Data Hub Center of the Korea Institute of Science and Technology Information
and
KREONET/GLORIAD;
Universiti Malaya RU grant, Akademi Sains Malaysia, and Ministry of Education Malaysia;
Frontiers of Science Program Contracts
No.~FOINS-296,
No.~CB-221329,
No.~CB-236394,
No.~CB-254409,
and
No.~CB-180023, and No.~SEP-CINVESTAV research Grant No.~237 (Mexico);
the Polish Ministry of Science and Higher Education and the National Science Center;
the Ministry of Science and Higher Education of the Russian Federation,
Agreement No.~14.W03.31.0026, and
the HSE University Basic Research Program, Moscow;
University of Tabuk research Grants
No.~S-0256-1438 and No.~S-0280-1439 (Saudi Arabia);
Slovenian Research Agency and research Grants
No.~J1-9124
and
No.~P1-0135;
Agencia Estatal de Investigacion, Spain
Grant No.~RYC2020-029875-I
and
Generalitat Valenciana, Spain
Grant No.~CIDEGENT/2018/020
Ministry of Science and Technology and research Grants
No.~MOST106-2112-M-002-005-MY3
and
No.~MOST107-2119-M-002-035-MY3,
and the Ministry of Education (Taiwan);
Thailand Center of Excellence in Physics;
TUBITAK ULAKBIM (Turkey);
National Research Foundation of Ukraine, project No.~2020.02/0257,
and
Ministry of Education and Science of Ukraine;
the U.S. National Science Foundation and research Grants
No.~PHY-1913789 
and
No.~PHY-2111604, 
and the U.S. Department of Energy and research Awards
No.~DE-AC06-76RLO1830, 
No.~DE-SC0007983, 
No.~DE-SC0009824, 
No.~DE-SC0009973, 
No.~DE-SC0010007, 
No.~DE-SC0010073, 
No.~DE-SC0010118, 
No.~DE-SC0010504, 
No.~DE-SC0011784, 
No.~DE-SC0012704, 
No.~DE-SC0019230, 
No.~DE-SC0021274, 
No.~DE-SC0022350, 
No.~DE-SC0023470; 
and
the Vietnam Academy of Science and Technology (VAST) under Grant No.~DL0000.05/21-23.

These acknowledgements are not to be interpreted as an endorsement of any statement made
by any of our institutes, funding agencies, governments, or their representatives.

\bibliographystyle{apsrev}
\bibliography{references}

\begin{thebibliography}{31}
\bibitem[{cha()}]{charge-conjugates}
\bibinfo{note}{Throughout this paper, charge-conjugate modes are implicitly
  included unless noted otherwise.}
\end{thebibliography}

\end{document}